\documentclass[sigconf,natbib=true]{acmart}
\usepackage{multicol}
\usepackage{multirow}
\usepackage{amsfonts}
\usepackage{bm}
\usepackage{amsmath}
\usepackage{amsthm}

\AtBeginDocument{%
  \providecommand\BibTeX{{%
    \normalfont B\kern-0.5em{\scshape i\kern-0.25em b}\kern-0.8em\TeX}}}

\begin{document}
\copyrightyear{2024} 
\acmYear{2024} 
\setcopyright{acmlicensed}\acmConference[WWW '24 Companion]{Companion Proceedings of the ACM Web Conference 2024}{May 13--17, 2024}{Singapore, Singapore}
\acmBooktitle{Companion Proceedings of the ACM Web Conference 2024 (WWW '24 Companion), May 13--17, 2024, Singapore, Singapore}
\acmDOI{10.1145/3589335.3651940}
\acmISBN{979-8-4007-0172-6/24/05}
\title{ConvSDG: Session Data Generation for Conversational Search}

\author{Fengran Mo}
\orcid{0000-0002-0838-6994}
\affiliation{%
  \institution{RALI, Université de Montréal}
  \city{Montréal}
  \state{Québec}
  \country{Canada}
}
\email{fengran.mo@umontreal.ca}

\author{Bole Yi}
\orcid{0009-0005-0898-1119}
\affiliation{%
  \institution{RALI, Université de Montréal}
  \city{Montréal}
  \state{Québec}
  \country{Canada}
}
\email{bole.yi@umontreal.ca}

\author{Kelong Mao}
\orcid{0000-0002-5648-568X}
\affiliation{%
  \institution{Renmin University of China} 
  \city{Beijing}
  \country{China}}
\email{mkl@ruc.edu.cn}

\author{Chen Qu}
\orcid{0000-0002-3273-7109}
\affiliation{%
  \institution{University of Massachusetts Amherst \city{Amherst}
  \state{MA}
  \country{USA}}
}
\email{mail@cqu.org}

\author{Kaiyu Huang}
\orcid{0000-0001-6779-1810}
\affiliation{%
  \institution{Beijing Jiaotong University}
  \city{Beijing}
  \country{China}}
\email{kyhuang@bjtu.edu.cn}

\author{Jian-Yun Nie}
\orcid{0000-0003-1556-3335}
\authornote{Corresponding author}
\affiliation{%
  \institution{RALI, Université de Montréal}
  \city{Montréal}
  \state{Québec}
  \country{Canada}
}
\email{nie@iro.umontreal.ca}

\renewcommand{\shortauthors}{Fengran Mo et al.}

\begin{abstract}
Conversational search provides a more convenient interface for users to search by allowing multi-turn interaction with the search engine. However, the effectiveness of the conversational dense retrieval methods is limited by the scarcity of training data required for their fine-tuning.
Thus, generating more training conversational sessions with relevant labels could potentially improve search performance. 
Based on the promising capabilities of large language models (LLMs) on text generation, we propose \textit{ConvSDG}, a simple yet effective framework to explore the feasibility of boosting conversational search by using LLM for session data generation. Within this framework, we design dialogue/session-level and query-level data generation with unsupervised and semi-supervised learning, according to the availability of relevance judgments. The generated data are used to fine-tune the conversational dense retriever.
Extensive experiments on four widely used datasets demonstrate the effectiveness and broad applicability of our ConvSDG framework compared with several strong baselines.
\end{abstract}




\maketitle

\section{Introduction}

Conversational search is an emerging area within information retrieval that is poised to become the future of search engines~\cite{gao2022neural}. The systems empower users to engage in interactive, multi-turn conversations when searching for information, simplifying the process of addressing intricate information needs. The central hurdle lies in accurately understanding users' genuine search intent, given that their queries are context-dependent and prone to linguistic issues like omission, coreference, and ambiguity~\cite{trec_cast20}.

When tackling conversational search, an intuitive approach is to use conversational query rewriting (CQR). This method involves breaking down the task by employing rewrite models to convert the query of the current turn into a de-contextualized one and then conducting an ad-hoc search using the rewritten query. This approach allows for the use of existing retrievers for the search process. However, it is challenging to directly optimize the rewriting towards search~\cite{lin2021contextualized,wu2022conqrr,mo2023convgqr,mao2023search}.
Another approach, known as conversational dense retrieval (CDR), focuses on training a conversational dense retriever to grasp the search intent by implicitly learning the latent representations of encoded queries and passages. Unlike CQR, the conversational dense retriever has the ability to naturally learn from the relevance signals between queries and passages.

Nonetheless, the current CQR and CDR techniques, which are trained on limited data, struggle to deliver satisfactory performance due to the prevalence of the long-tail phenomenon~\cite{mao2022COTED,mo2023learn} within conversational sessions and the scarcity of available samples in existing datasets~\cite{trec_cast19,trec_cast20,trec_cast21}. Creating conversational search datasets manually is a costly and difficult endeavor, so an intuitive approach is to automatically enrich the session data for model training. While some prior studies~\cite{mao2022convtrans,dai2022inpainting} have demonstrated its feasibility, the additional session data generated in this manner often lack the necessary power for continuous model improvement, and the generation process remains complex. Furthermore, these approaches still rely on the assumption that there is a substantial amount of relevant in-domain data available to build data augmentation models.
The recent success of large language models (LLMs)~\cite{neurips20_GPT-3,iclr22_FLAN,arxiv22_LaMDA,zhang2023moqagpt}, which excel in generating texts, has brought notable advancements to the field of information retrieval~\cite{mao2020item}. These LLMs are now being applied to support various techniques within the field, such as query expansion~\cite{arxiv23_query2doc}, query generation~\cite{arxiv22_InPars,ICLR23_Promptagator}, and document prediction~\cite{arxiv22_HyDE,arxiv23_GRF}.
Motivated by these developments, our research explores the potential of harnessing LLMs for the automatic generation of session data, adapted appropriately to enhance conversational search performance. In essence, we aim to address the following research questions related to the utilization of LLMs:

\textbf{RQ1}: Is it feasible to exploit automatic session data generation for conversational search models to achieve comparable or better performance than relying on manually constructed datasets?

\textbf{RQ2}: Can we improve conversational search performance by augmenting the diversity of session queries via query rewriting and existing annotations?


In order to address these inquiries, we introduce \textit{ConvSDG}, a data augmentation framework aimed at employing Large Language Models (LLMs) to accomplish \textbf{S}ession \textbf{D}ata \textbf{G}eneration for \textbf{Conv}ersational search. Our framework leverages the robust text generation capabilities of LLMs to automatically generate session data that can adapt effectively to the demands of conversational search scenarios. With well-defined supervision signals, this approach mitigates the problem of limited data availability and enhances the performance of conversational dense retrieval.
Specifically, we designed two different prompts in our framework, each tailored to specific scenarios, enabling LLMs to generate dialogue-level session data and query-level augmented session data. Subsequently, we create or assign appropriate supervision signals for each query turn, catering to both unsupervised and semi-supervised learning settings. Finally, the generated session data, along with the annotations, are used to fine-tune the conversational dense retriever. We carry out comprehensive experiments using four widely used conversational search datasets and compare ConvSDG against several strong baselines, demonstrating its superior performance.

Our contributions are summarized as follows:

(1) We propose a simple yet effective framework to automatically generate session data for conversational search, showing the feasibility of using automatic data to train the models.

(2) We develop two approaches for instructing the LLM to produce session data, at both dialogue and query levels. Additionally, we generate the necessary supervision signals to facilitate conversational dense retrieval with different learning manners.

(3) We demonstrate the effectiveness of ConvSDG by achieving better results compared to models trained on manually curated data across four datasets and under two distinct settings. The analysis offers additional insights into the potential of automatic data generation to enhance conversational search.

\section{Related Work}
\subsection{Conversational Search}
Conversational query rewriting (CQR) and conversational dense retrieval (CDR) represent the two primary approaches to conversational search. CQR focuses on transforming context-dependent queries within a search session into stand-alone queries. Common methods involve selecting relevant tokens from the search session~\cite{voskarides2020query,lin2021HQE,qian2022explicit} and training a generative rewriter model using human-rewritten queries paired with their respective sessions~\cite{lin2020conversational,yu2020few,vakulenko2021question,mao2023large}. Some research efforts incorporate reinforcement learning~\cite{wu2022conqrr,chen2022RLCQR} or ranking signals~\cite{mo2023convgqr,mao2023search} to align the generation process with the downstream search task. In contrast, CDR utilizes conversational search session data to perform end-to-end dense retrieval. To enhance conversational search performance, advanced techniques like context denoising~\cite{qu2020open,yu2021few,krasakis2022zeco,mao2022COTED,mao2023learning,mo2023learn}, data augmentation~\cite{lin2021contextualized,dai2022inpainting,mao2022convtrans}, and the mining of challenging negative examples~\cite{kim2022saving,mo2024history} have been explored.

\subsection{Large Language Models for Data Generation}
The quantity and quality of data hold significant value across various research domains within natural language processing (NLP) and information retrieval (IR). The emergence of pre-trained language models~\cite{devlin2019bert,radford2019language}, and more recently, large language models~\cite{neurips20_GPT-3,iclr22_FLAN,arxiv22_LaMDA}, has opened up new opportunities for automatic text generation. For example, some studies utilize these language models to generate data for a wide array of NLP tasks, including text classification~\cite{wang2021towards}, acquiring commonsense knowledge~\cite{west2022symbolic}, natural language inference~\cite{liu2022wanli}, open-domain dialogue generation~\cite{zheng2023augesc}, and sequence labeling~\cite{ding2023gpt}. The enhanced performance observed in downstream tasks through these approaches validates the effectiveness of employing language models for data generation. 

\subsection{LLM-based Data Generation for IR}
There are also several other approaches using large language models (LLMs) to generate data for ad-hoc IR: to generate queries from a document~\cite{arxiv22_InPars,ICLR23_Promptagator}, to generate a document from a query~\cite{arxiv22_HyDE,arxiv23_GRF,arxiv23_query2doc}, etc. 
Different from them, our work concentrates on investigating how LLMs can be harnessed to create conversational search data, an underexplored area in the existing literature. A recent work -- CONVERSER~\cite{huang2023converser} focuses on few-shot query generation with in-context learning. It relies on two-stage generation and the needed generated samples are quite large, while we only require one-stage generation with much less generated samples (higher efficiency). Besides, this work is narrower in its scope compared to ours, and its effectiveness, evaluated on the CAsT-19 dataset only, is much lower than our method and other state-of-the-art baselines.

\section{Methodology}
\subsection{Task Definition}
Conversational search 
tries to identify relevant passages $p^*$ from a vast collection $\mathcal{C}$ in response to the current query ($n$-th) $q_n$. This process is based on the context provided by the ongoing conversation session $\mathcal{S}=\{q_i\}_{i=1}^{n-1}$. Each query turn within a session depends on the preceding context, necessitating the conversational retriever to possess the capability to comprehend the user's search intent. Thus, having access to high-quality and adequate conversational search session data is important.
The goal of this paper is to generate new session data $\mathcal{S^{\prime}}=\{q_i^{\prime}\}_{i=1}^{n}$ based on LLMs and produce a series of query-passage pairs $\{(q_i^{\prime}, p_i^{\prime})\}_{i=1}^{n}$ for fine-tuning conversational dense retrieval. 

\begin{figure*}[t]
\centering
\includegraphics[width=0.9\textwidth]{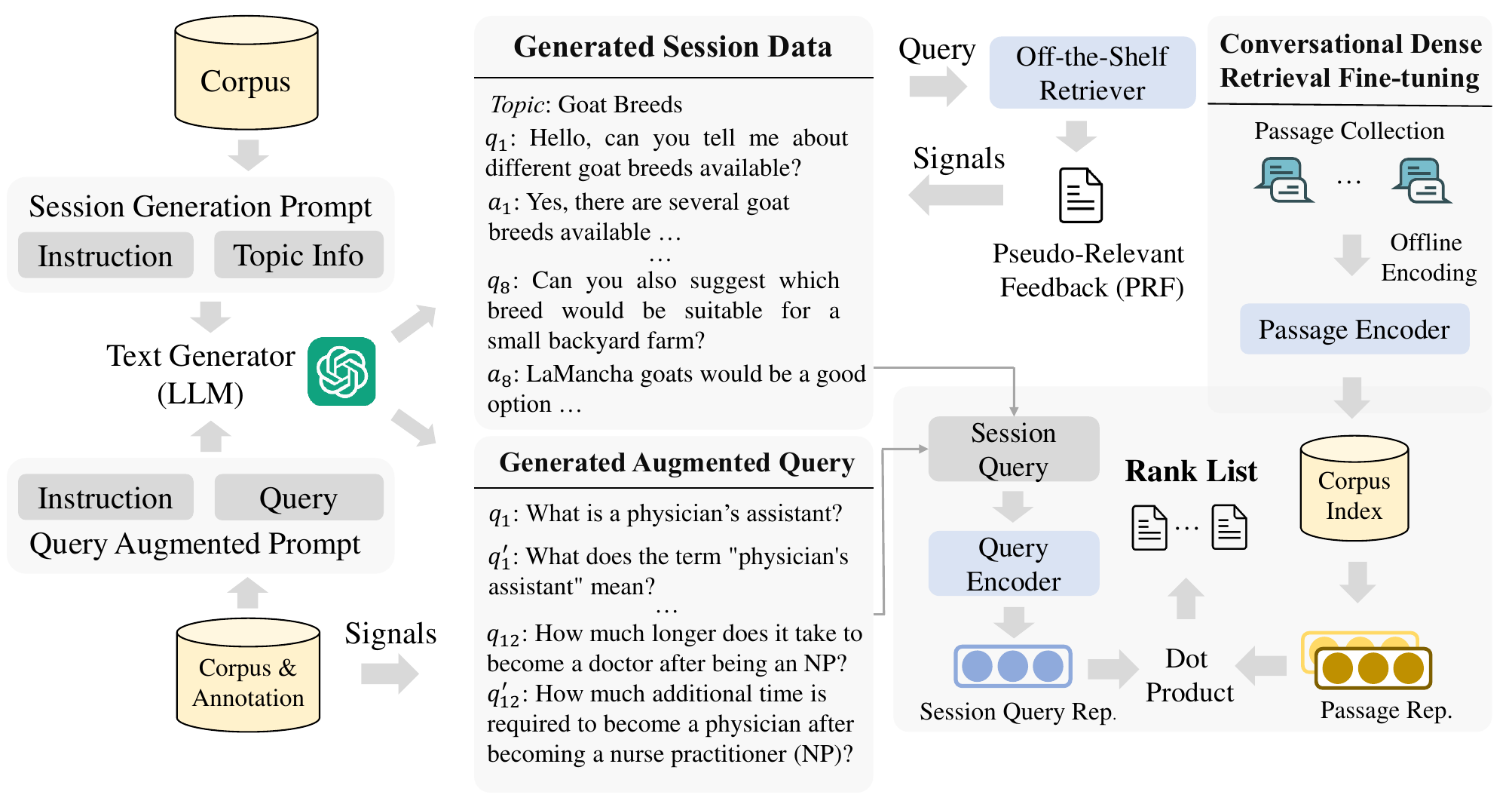}
\caption{Overview of ConvSDG. Three parts are included: (1) Two prompts for session data generation at different levels, (2) Produce supervision signals for generated data, PRF for session generation and existing annotations for query augmented, (3) Conduct conversational dense retrieval fine-tuning with the generated data.}
\label{fig: overview}
\end{figure*}

\subsection{Overview}
The construction of existing conversational search datasets~\cite{trec_cast19,trec_cast20,trec_cast21} heavily relies on human effort, resulting in insufficient data samples to adequately support fine-tuning of end-to-end conversational dense retrievers.
Drawing inspiration from recent successes in harnessing Large Language Models (LLMs) for data generation in various downstream tasks, we introduce the \textit{ConvSDG} framework. This framework explores the potent generative capabilities of LLMs to create high-quality conversational session data for conversational search, regardless of whether relevant judgments are available or not.
In the case where relevant judgments are unavailable, we employ LLMs to efficiently generate the entire conversational sessions at the dialogue-level, using only the topic description. We then apply pseudo-relevance feedback as supervision signals. In contrast, when relevant judgments exist in the dataset, we perform query-level augmentation by rephrasing the query formulation for each turn, recognizing that queries with the same search intent can be expressed in multiple ways.

The overview of our ConvSDG is depicted in Fig.~\ref{fig: overview}. It consists of three main steps: (1) Guiding the LLM to generate session data at two different levels, (2) Generating supervision signals for each generated query turn, and (3) Conducting fine-tuning of conversational dense retriever based on the generated data.

\subsection{Dialogue-level Session Generation} 
\label{sec: Dialogue-level Session Generation}
A conversational session typically centers around a particular topic~\cite{adlakha2022topiocqa}, like ``animals'', and each turn explores different aspects of that topic, such as ``habits'' or ``various breeds''. To mimic the real-world scenario, it is essential to consider specific conversational phenomena, such as ensuring coherent transitions between turns, handling co-references, and addressing instances of omission, when constructing these conversational sessions~\cite{zamani2023conversational}. Based on our initial experiments and insights gleaned from existing research~\cite{kim2022botstalk,zheng2023augesc}, we have found that generating a conversation session one turn at a time using generative models does not yield high-quality results. Additionally, we have noticed that maintaining turn-to-turn coherence for Large Language Models (LLMs) solely through prompt instructions is challenging. As a solution, we opt for dialogue-level session data generation, which involves creating the entire conversation session in one go by providing a specific topic description. This approach helps avoid the generation of inconsistent query turns.

\begin{figure*}[t]
\centering
\includegraphics[width=0.9\textwidth]{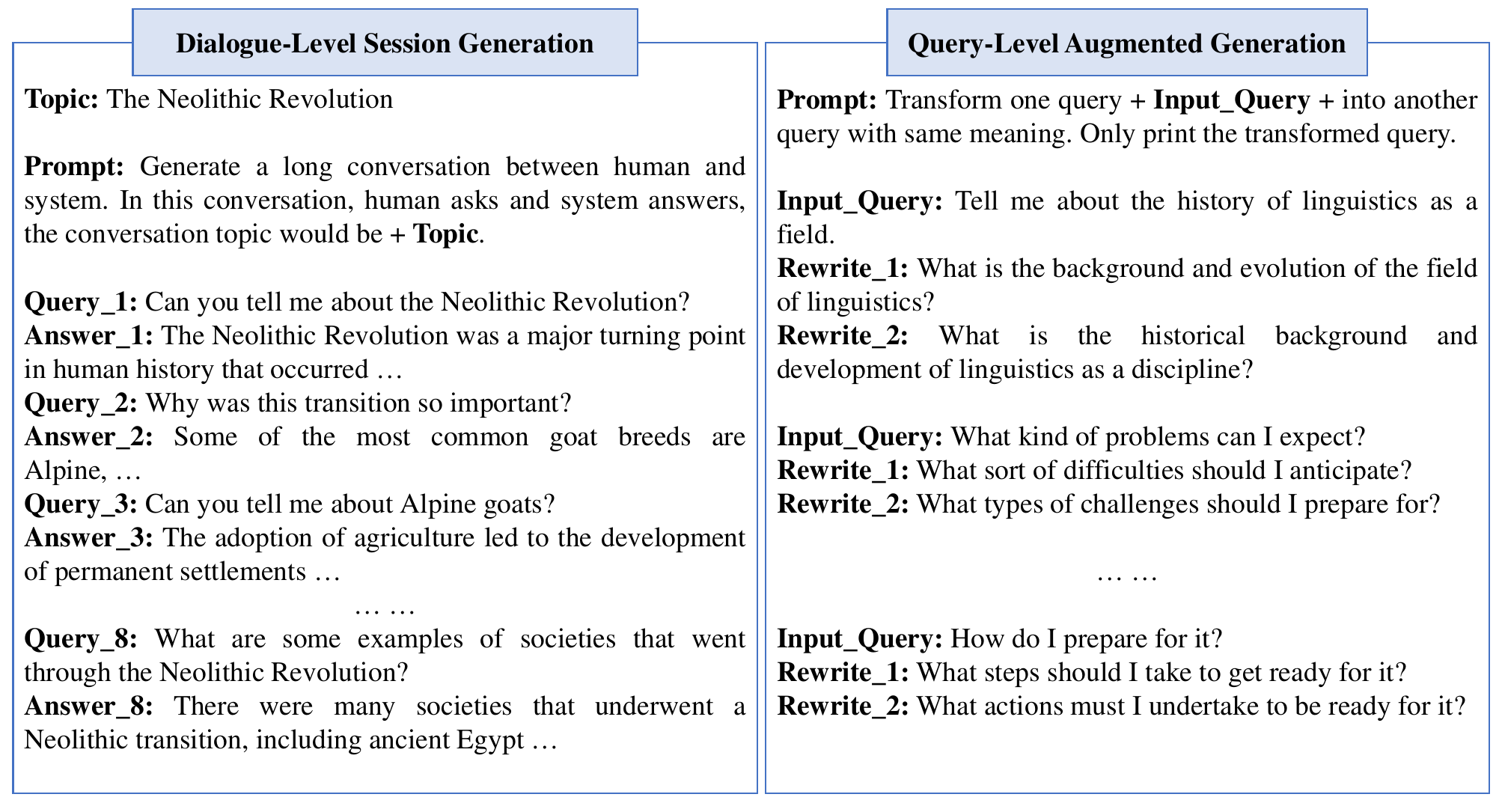}
\caption{An example to illustrate the conversational session data generation for both dialogue-level (left) and query-level (right).}
\label{fig: prompt}
\end{figure*}

In detail, we begin by sampling the topic description for a session from existing datasets\footnote{Topic descriptions can be sampled in any suitable ways.}, which we then use to create a prompt instruction. Our prompt instruction template is structured as \textit{[Instruction, Topic Information]}, enabling the LLM to generate an entire session in one go. A comprehensive illustration is shown in Fig.~\ref{fig: prompt}. 

Once we have the generated session data, we require relevance judgments that establish the connection between query turns and passages for the fine-tuning of the conversational dense retriever. In practice, obtaining such annotations is significantly more costly compared to acquiring conversation session data~\cite{trec_cast19}. To enhance the flexibility of our framework, we attempt to train models in an \textbf{unsupervised} manner, i.e. we do not rely on relevance judgments provided by human experts. Instead, we adopt the idea of pseudo-relevance feedback~\cite{tao2007exploration} to create 
pseudo supervision signals for each query turn. It is worth noting that there is not a single fixed method for this purpose, and we leave further exploration of this area for future research.

Concretely, we perform off-the-shelf retrieval on each query turn, selecting three passages from top-5 at random as pseudo-relevant documents 
for that specific turn. 
However, it is important to consider the format of the input query used for this off-the-shelf retrieval because the original queries in the conversational session are not stand-alone and always rely on the context of the ongoing conversation.
To prevent topic shifts and find an optimal solution, we explore four query forms that take contextual information into account: (1) \textit{q+a}: which concatenates the current query and the corresponding hypothetical answer generated by LLM, (2) \textit{q+a+topic}: which combines the current query, its answer, and the session topic information, (3) \textit{convq+topic}: which concatenates the current query, all previous queries, and the session topic information, (4) \textit{convq+conva+topic}: which combines the current query, all previous queries and their corresponding answers, along with the session topic information. Ultimately, we select the \textit{q+a+topic} format for reformulating the input query due to its demonstrated effectiveness, as discussed later in Sec.~\ref{sec: Query Form}.
 
\subsection{Query-level Augmented Generation}
While utilizing pseudo-relevance feedback to create supervision signals is efficient and often yields comparable results to fully supervised methods, there is still a potential downside of introducing false positive signals that could misguide model training. To mitigate this risk, we can leverage the limited relevance judgments provided by human experts in existing conversational search datasets to carry out query-level augmented generation. Our objective is to generate additional conversational search session data based on the original annotations, specifically for each query turn. The underlying assumption here is that the sequence of conversation should not be unique. In other words, the same user search intent can be expressed in different natural language forms, leading to various conversational sessions on the same topic. This variability is common in real-world scenarios.

Our method operates on the principle of generating new data by making adjustments to existing data points, guided by prior knowledge about the problems' underlying structure. The augmented data points, derived from labeled data within our framework, can be directly applied in semi-supervised learning through consistency regularization. 
``Semi-supervised'' means 
we combine the original data points with manual labels and the generated data points without manual labels for model training.
In practice, we prompt the LLM to rewrite each query, providing an alternative natural language expression with the same meaning. This instruction template adheres to the format \textit{[Instruction, Input Query]}, as illustrated in Fig.~\ref{fig: prompt}. By repeating this process $t$ times, we generate $t$ augmented queries for each original query turn, effectively expanding the initial dataset by a factor of $t$. Subsequently, the original relevance judgments for each query turn are associated with each corresponding augmented query, serving as supervision signals.


\subsection{Conversational Dense Retrieval}
We conduct fine-tuning for conversational dense retrieval using the session data we have generated and the associated supervision signals. For this task, we employ a widely used ad-hoc search retriever, ANCE~\cite{xiong2020approximate}, which serves as both the query and passage encoder, denoted as $\mathcal{F}_Q$ and $\mathcal{F}_P$, respectively. In this process, we consider all preceding queries within the same session to reformulate the current query turn $q_n^{ref}$, expressed as
\begin{equation}
    q_n^{ref} = q_1 \circ \cdots q_{i} \cdots \circ q_{n-1} \circ q_n, \quad i \in [1, n-1]
    \label{eq: query form}
\end{equation}
where $\circ$ denotes concatenation. Then, a similarity function $\mathcal{S}$ based on dot product is applied to score a candidate passage $p$ as:
\begin{equation}
    \mathcal{S}(q_n^{ref}, p) = \mathcal{F}_{Q}(q_n^{ref})^{T} \cdot \mathcal{F}_{P}(p)
    \label{eq: similarity}
\end{equation}
During the training phase, we update only the query encoder, while the passage encoder is frozen. The training objective follows the widely used contrastive learning loss:
\begin{equation}
\mathcal{L} = \frac{e^{\mathcal{S}(q_n^{ref}, p^+)}} {e^{\mathcal{S}(q_n^{ref}, p^+)} + \sum_{p_n^- \in C^-} e^{\mathcal{S}(q_n^{ref}, p^-)}}
    \label{eq: loss}
\end{equation}
where the $p^+$ and $p^-$ denote the positive and negative samples for each query turn. During the inference phase, we perform the Approximate Nearest Neighbors (ANN) search based on the dense index using Faiss~\cite{johnson2019billion}.

\section{Experimental Settings}
\subsection{Datasets and Evaluation Metrics}
We carry out extensive experiments on four widely used conversational search datasets: \textbf{CAsT-19}~\cite{trec_cast19}, \textbf{CAsT-20}~\cite{trec_cast20}, \textbf{CAsT-21}~\cite{trec_cast21}, and \textbf{TopiOCQA}~\cite{adlakha2022topiocqa}.
The three CAsT datasets are curated by the experts of the TREC Conversational Assistance Track (CAsT) and each dataset comprises information-seeking conversations encompassing hundreds of turns in total.
The TopiOCQA dataset addresses the novel challenge of topic switching, a common phenomenon in real-world scenarios. 
Table~\ref{table:data_statistics} provides an overview of the fundamental statistics of the datasets.\footnote{These statistics consider only the turns that possess relevance labels.} Following previous studies~\cite{trec_cast20,trec_cast21,krasakis2022zeco,mao2023large}, we employ Mean Reciprocal Rank (MRR), NDCG, and Recall as our evaluation metrics, computed with the \texttt{pytrec\_eval} tool~\cite{sigir18_pytrec_eval}. 

\subsection{Baseline Models}
We compare ConvSDG with the following models:
(1) \textbf{BM25}~\cite{robertson2009probabilistic}: A widely used unsupervised sparse retrieval model.
(2) \textbf{ConvDR}~\cite{yu2021few}: A conversational dense retrieval model based on ANCE retriever, containing both zero-shot and few-shot manner, which is supervised by both conversational search data and manual rewritten queries.
(3) \textbf{ZeCo}~\cite{krasakis2022zeco}: A zero-shot conversational search method that matches only the contextualized terms of the current query with passages based on ColBERT~\cite{sigir20_colbert}.
(4) \textbf{LLMQR}: A conversational query rewriting method based on the large language model ChatGPT (gpt-turbo-3.5-4k) without supervision signals to directly rewrite the current turn with the given conversation session.
(5) \textbf{CONVERSER}~\cite{huang2023converser}: A few-shot two-stage conversational query generation method based on the large language model for training conversational dense retrievers.
(6) \textbf{T5QR}~\cite{lin2020conversational}: A conversational query rewriting method based on the T5 model using human-rewritten queries in the QReCC dataset~\cite{anantha2021open}. 
(7) \textbf{ConvGQR}~\cite{mo2023convgqr}: A conversational query reformulation method by combining query rewrite and expansion, which leverages human-rewritten queries and gold answers in QReCC dataset as generation objectives.
(8) \textbf{w/o Aug.}~\cite{mao2023learning}: A conversational dense retriever that fine-tunes ANCE with the original (non-augmented) conversational search data. We use it as the baseline of our methods to see the impact of data augmentation. The fine-tuning process is only based on the CAsT-19 training set. 

\begin{table}[!t]
\centering
\caption{Statistics of the three CAsT and TopiOCQA datasets.}
\scalebox{0.85}{
\begin{tabular}{lcccccc}
\toprule
\multirow{2}{*}{\textbf{Dataset}} & \multicolumn{2}{c}{\textbf{CAsT-19}}  & \textbf{CAsT-20}    & \textbf{CAsT-21}  & \multicolumn{2}{c}{\textbf{TopiOCQA}} \\ 
\cmidrule(lr){2-3}\cmidrule(lr){4-4}\cmidrule(lr){5-5}\cmidrule(lr){6-7}
     & \textbf{Train} &  \textbf{Test} & \textbf{Test} &  \textbf{Test} & \textbf{Train} & \textbf{Test} \\
\midrule
\# Conversations  & 30 &  20  & 25         & 18 & 3,509 & 205  \\
\# Turns (Queries) & 108 & 173 & 208        & 157  &  45,450 & 2,514 \\
\# Passages/Docs    & \multicolumn{2}{c}{38M} & 38M & 40M & \multicolumn{2}{c}{25M}  \\ \bottomrule
\end{tabular}}
\label{table:data_statistics}
\end{table}

\subsection{Implementation Details}
We utilize OpenAI's ChatGPT (gpt-turbo-3.5-4k) API for both dialogue-level session generation and query-level augmented generation with the default hyper-parameters. For the pseudo relevance supervision signals, we randomly select three passages from the top-5 retrieved passages using PRF. 
The ANCE system serves as the backbone model for fine-tuning conversational dense retrieval, with maximum input lengths truncated at 64 for queries, 384 for passages, and 512 for concatenated sessions. Model training employs a batch size of 16 with 5 epochs. Further details can be found in our released code repository\footnote{\url{https://github.com/fengranMark/ConvSDG}}.

\subsection{Experimental Scenarios}
We evaluate our method on the following two training scenarios, with and without the relevance judgment, and compare with the suitable baseline models: \\

\noindent \textbf{Dialogue-level augmentation w/o relevance judgment:}
We utilize solely the topic information from the CAsT-19 and TopiOCQA training sets for ConvSDG to perform dialogue-level session generation. Subsequently, we assess its performance on the respective test sets of all three CAsT datasets and TopiOCQA. Consequently, in the absence of existing relevance judgments, conversational dense retrieval fine-tuning is carried out following an unsupervised learning approach. The methods we compare include unsupervised and zero-shot methods, as well as the direct use of LLM. \\

\noindent \textbf{Query-level augmentation w/ relevance judgment:}
We employ ConvSDG for query-level augmented generation and use the limited relevance judgments available in the CAsT-19 training set. The evaluation is then carried out on three CAsT datasets. As a result, conversational dense retrieval fine-tuning takes place in a semi-supervised learning manner, with the compared methods being supervised and trained using the same data samples.


\section{Experimental Results}

\begin{table*}[t]
    \centering
    \caption{Performance of two different settings on CAsT datasets. $\dagger$ denotes significant improvements with t-test at $p<0.05$ over all compared methods and \textbf{bold} indicates the best results in corresponding settings. The turns of CONVERSER are quoted from the original paper and the turns of our ConvSDG with relevance judgment are expanded two times and combined with the original 745 turns in the original training set.}

    \begin{tabular}{c|c|ccc|ccc|ccc}
    \toprule
        \multirow{2}{*}{Method}  & \multirow{2}{*}{Turn} & \multicolumn{3}{c|}{CAsT-19} & \multicolumn{3}{c|}{CAsT-20} & \multicolumn{3}{c}{CAsT-21}\\ \cmidrule{3-11}
        ~ & ~ & MRR & NDCG@3 & Recall@100 & MRR & NDCG@3 & Recall@100 & MRR & NDCG@3 & Recall@100\\ 
        \midrule
        \multicolumn{11}{c}{Dialogue-level augmentation w/o relevance judgement}\\
        \midrule
        BM25 & - & 39.7 & 18.0 & 20.1 & 13.9 & 7.2 & 11.5 & 30.3 & 16.6 & 24.9\\
        ZeCo & - & - & 23.8 & 21.6 & - & 17.6 & 20.0 & - & 23.4 & 26.7\\
        ConvDR & - & 42.0 & 24.7 & 18.3 & 23.4 & 15.0 & 15.0 & - & - & -\\
        LLMQR & - & 57.8 & \textbf{35.0} & 27.7 & 36.8 & 24.5 & 28.2 & 42.1 & 28.2 & 31.3\\
        CONVERSER & 230k & 35.8 & 21.4 & - & - & - & - & - & - & - \\
        \midrule
        ConvSDG (Ours) & 1,704 & \textbf{59.5}$^\dagger$ & 32.1 & \textbf{33.5}$^\dagger$ & \textbf{37.9}$^\dagger$ & \textbf{25.3}$^\dagger$ & \textbf{34.9}$^\dagger$ & \textbf{50.2}$^\dagger$ & \textbf{33.2}$^\dagger$ & \textbf{37.5}$^\dagger$\\
        \bottomrule
        \bottomrule
        \multicolumn{11}{c}{Query-level augmentation w/ relevance judgement}\\
        \midrule
        T5QR & 2,235 & 52.8 & 31.3 & 25.3 & 29.7 & 19.0 & 24.2 & 30.3 & 20.5 & 20.6\\
        ConvGQR & 2,235 & 61.0 & 34.6 & 30.3 & 35.1 & \textbf{24.3} & 30.3 & 37.6 & 24.6 & 28.4\\
        ConvDR & 2,235 & \textbf{62.1} & 35.0 & 29.7 & 34.6 & 23.6 & 28.8 & 37.6 & 25.2 & 31.4\\
        \midrule
        w/o Aug. & 745 & 56.8 & 31.5 & 29.2 & 34.2 & 22.6 & 32.6 & 45.6 & 29.8 & 35.2\\
        ConvSDG (Ours) & 2,235 & 60.6 & \textbf{35.3} & \textbf{32.1}$^\dagger$ & \textbf{36.5}$^\dagger$ & 24.2 & \textbf{34.2}$^\dagger$ & \textbf{47.2}$^\dagger$ & \textbf{30.8}$^\dagger$ & \textbf{36.8}$^\dagger$\\
        \bottomrule
     \end{tabular}
     \label{table: CAsT results}
\end{table*}

\begin{table*}[t]
    \centering
    \caption{Performance on TopiOCQA dataset. $\dagger$ denotes significant improvements with t-test at $p<0.05$ over all compared methods and \textbf{bold} indicates the best results (except for the results for reference).}
    \begin{tabular}{c|c|cccc}
    \toprule
        \multirow{2}{*}{Method}  & \multirow{2}{*}{Turn} & \multicolumn{4}{c}{TopiOCQA} \\ 
        \cmidrule{3-6}
        ~ & ~ & MRR & NDCG@3 & Recall@10 & Recall@100 \\ 
        \midrule
        \multicolumn{6}{c}{Dialogue-level augmentation w/o relevance judgement}\\
        \midrule
        BM25 & - & 10.7 & 9.7 & 11.2 & 26.7 \\
        ConvDR & - & 10.3 & 9.1 & 15.7 & 35.4\\
        \midrule
        ConvSDG (Ours) & 5,231 & \textbf{21.4}$^\dagger$ & \textbf{19.9}$^\dagger$ & \textbf{37.8}$^\dagger$ & \textbf{58.0}$^\dagger$ \\
        \midrule
        \multicolumn{6}{c}{Query-level augmentation w/ relevance judgement}\\
        \midrule
        T5QR & 5,231 & 18.4 & 17.6 & 30.8 & 45.3 \\
        ConvGQR & 5,231 & 8.0 & 7.3 & 14.3 & 25.5 \\
        \midrule
        \multicolumn{6}{c}{For Reference}\\
        \midrule
        T5QR & 63,501 & 23.0 & 22.2 & 37.6 & 54.4 \\
        ConvGQR & 63,501 & 25.6 & 24.3 & 41.8 & 58.8 \\
        \bottomrule
     \end{tabular}
     \label{table: TopiOCQA results}
\end{table*}

\subsection{Main Results}
The overall performance comparisons on CAsT and TopiOCQA datasets with different settings are presented in Table~\ref{table: CAsT results} and Table~\ref{table: TopiOCQA results}.  

\begin{figure*}[t]
\centering
\includegraphics[width=0.65\textwidth]{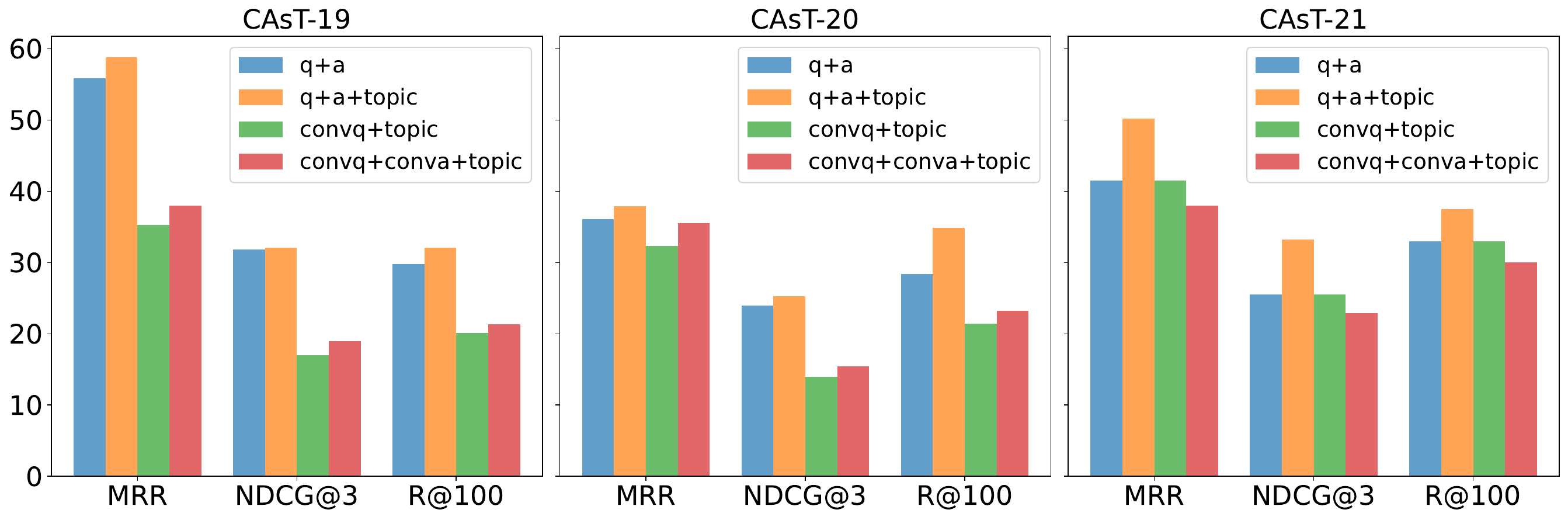}
\caption{Effectiveness of using generated supervision signals by different query forms based on ANCE dense retriever.}
\label{fig: query form}
\end{figure*}

\begin{figure}[t]
\centering
\includegraphics[width=0.49\textwidth]{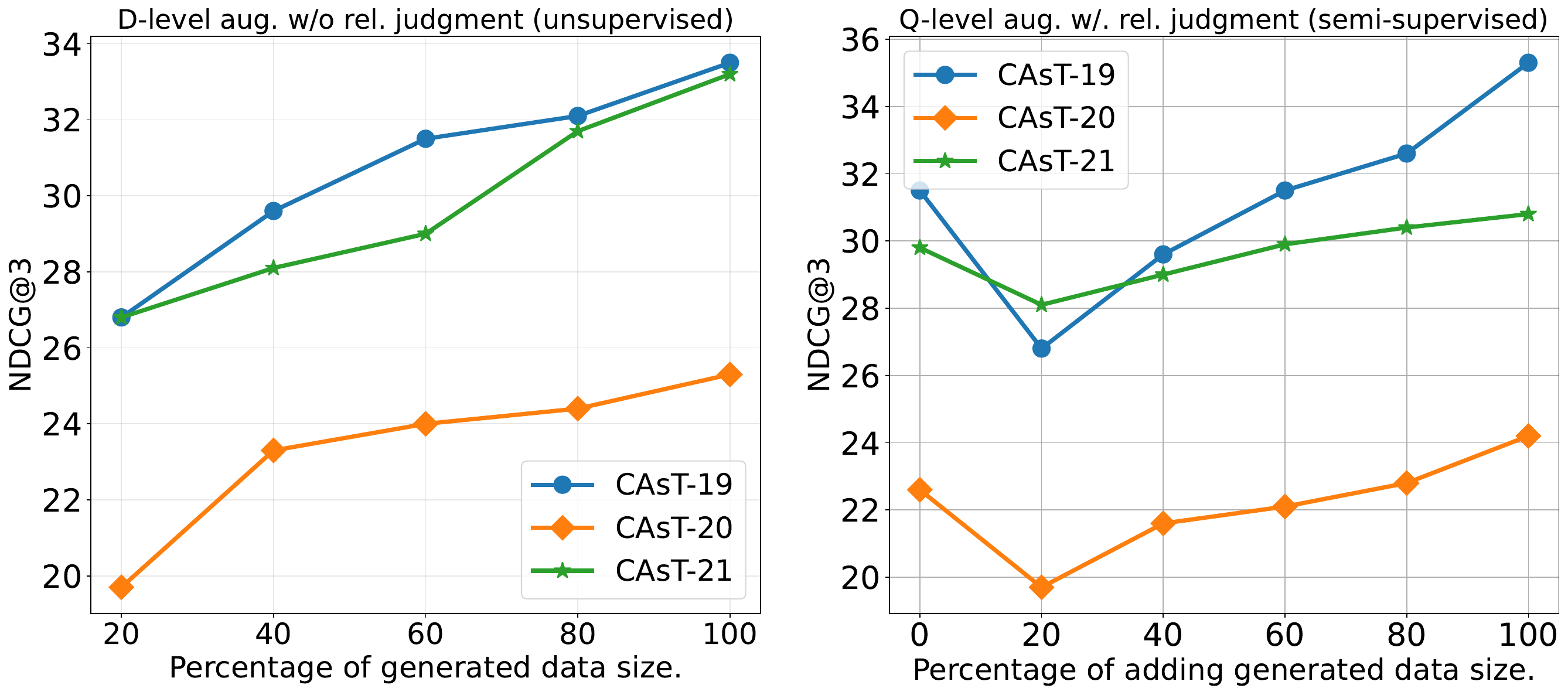}
\caption{Effectiveness of different sizes of generated data used for conversational fine-tuning with unsupervised (left) and semi-supervised (right) learning manner.}
\label{fig: data size}
\end{figure}

In the absence of relevance judgments with dialogue-level augmentation, as shown in Table~\ref{table: CAsT results}, we observe that ConvSDG, with unsupervised training, outperforms the compared methods across most evaluation metrics for both the CAsT and TopiOCQA datasets. In particular, it exhibits a significant relative improvement of 19.2\% in MRR and 17.7\% in NDCG@3 over the second-best results on the challenging CAsT-21 dataset. Besides, with fewer training samples, it even outperforms models trained with manual relevance judgments on CAsT-20 and CAsT-21. These findings confirm the quality and usefulness of our automatically generated data. 
Moreover, our approach outperforms all compared methods by applying the same augmented samples and is on par with supervised training methods with full original training samples on the TopiOCQA dataset (as shown in Table~\ref{table: TopiOCQA results}). These results emphasize the importance of conversational dense retrieval fine-tuning, especially when compared to zero-shot methods like ZeCo and ConvDR. Overall, our approach addresses the data scarcity challenge effectively through the automatic generation of conversational search data, validating 
our motivation.

When considering the scenario with query-level augmentation relevance judgments in the training data, as presented in Table~\ref{table: CAsT results}, ConvSDG continues to outperform the compared methods across most evaluation metrics with the same training samples on manual datasets. Specifically, it achieves substantial improvements by a relative boost of 25.5\% in MRR, 22.0\% in NDCG@3, and 17.2\% in Recall@100 over the second-best results on the more challenging CAsT-21 dataset. These enhancements, facilitated by the augmented query turns, show the effectiveness of our approach in rewriting the queries in the original datasets while retaining the same search intent. 
However, compared to dialogue-level ConvSDG without relevance judgment, ConvSDG with query-level augmentation is not more effective, even though it leverages human relevance judgments. 
This result suggests that there is still room for further improvement in system performance by fully harnessing the existing relevance annotations and enhancing the diversity of the conversational sessions.
It is also worth noting that the compared methods do not perform well on the more challenging CAsT-21 dataset. This discrepancy could be attributed to the fact that these methods depend on human-rewritten queries, and our generated augmented session queries may not align perfectly with these original annotations. This observation implies that CDR methods might be more suitable than CQR when being trained on the augmented queries.

\subsection{Effectiveness of Supervision Signals}
\label{sec: Query Form}

We present the retrieval performance achieved using four different query input forms, as discussed in Sec.~\ref{sec: Dialogue-level Session Generation}, for generating pseudo-relevance feedback (PRF) based on the ANCE dense retriever in Fig.~\ref{fig: query form}. In this setup, the top results obtained from PRF are employed as the pseudo supervision signals for the corresponding session query turns generated to fine-tune the conversational dense retriever.
Our findings indicate that using only the information from the current turn, i.e., the current turn's query and the corresponding hypothetical answer generated by LLM, the method tends to yield better results compared to incorporating the entire conversation context, such as concatenating with queries or answers from previous turns. This observation can be attributed to the fact that ad-hoc search retrievers lack the capability to represent an entire conversation session effectively, and the underlying search intent within the current turn query is context-dependent.
Moreover, we observe that the inclusion of topic information in each conversation session proves beneficial. Indeed, the topic information will help the generation process of augmented data to produce more relevant and topic-related data, which is better than that produced without the topic information.
Although topic information may not always be available during the inference stage, we can still leverage it to construct training datasets.




\subsection{Effectiveness of Generated Data Size}

We present an analysis of the effectiveness of employing varying sizes of generated data for conversational dense retrieval fine-tuning in two different scenarios across three CAsT datasets, as depicted in Fig.~\ref{fig: data size}. The percentage counted by the whole generated query turns for both unsupervised and semi-supervised settings.
For unsupervised learning, we observe a notable improvement in system performance as the volume of utilized data increases. This observation demonstrates the pivotal role of augmented data in mitigating the data scarcity issue, and it suggests that there is further potential for enhancing model performance by expanding the dataset even more.
On the other hand, for semi-supervised learning, we note that the fine-tuned models do not exhibit improved performance consistently compared to models trained solely on the original training set until a sufficient amount of generated data is added. This indicates that the generated data points might alter the data distribution, and it implies the need for appropriate filtering mechanisms. Nonetheless, the performance enhancement achieved with the addition of full-sized data underlines the effectiveness of our generated data for model training.

\section{Conclusion}
In this paper, we introduce \textit{ConvSDG}, a novel framework for generating session data with LLMs for conversational search. By harnessing the robust text generation capabilities of LLMs, we are able to fine-tune conversational dense retrieval using the session data generated through unsupervised or semi-supervised learning methods. Our experimental findings, based on four public datasets, demonstrate the remarkable effectiveness of our approach, as it outperforms existing comparable methods and even surpasses fully supervised models. Additionally, we analyze some crucial impacts of automatically constructing conversational search session data, offering insights for future exploration in this domain. Our study shows the effectiveness of using an LLM to generate additional training data for fine-tuning a dense retriever. It enriches the already extensive body of studies trying to exploit LLMs for search. More research is still required to find the best way to leverage LLMs for enhancing conversational search.


\bibliographystyle{ACM-Reference-Format}
\bibliography{sample-base}

\appendix


\end{document}